# Communication Dualism in Distributed Systems with Petri Net Interpretation

**Stanislaw Chrobot,**
**Wiktor B. Daszczuk**
Institute of Computer Science, Warsaw University of Technology
wbd@ii.pw.edu.pl

**Abstract:** In the paper notion of communication dualism id formalized and explained in Petri net interpretation. We consider communication dualism a basic property of communication in distributed systems. The formalization is done in the Integrated Model of Distributed Systems (IMDS) where synchronous communication, as well as asynchronous message-passing and variable-sharing are modeled in a common framework. In the light of this property, communication in distributed systems can be seen as a two-dimensional phenomenon with passing being its spatial dimension and sharing its temporal dimension. Any distributed system can be modeled as a composition of message-passes asynchronous processes or as a composition of variable-sharing asynchronous processes. A method of automatic process extraction in Petri net interpretation of IMDS is presented.

## 1 Introduction

The goal of this paper is a formalization of the notion of communication dualism. We show that passing and sharing exist in every distributed system, regardless of how it is designed. Communication dualism roughly corresponds to the idea of dualism first suggested informally in the famous Lauer-Needham hypothesis [16]. And even though not all conclusions of the hypothesis can be confirmed now, we consider communication dualism a basic property of distributed programming.

### 1.1 Integrated Model of Distributed Systems - IMDS

The formalization is done in the Integrated Model of Distributed Systems (IMDS) where synchronous communication, as well as asynchronous message-passing and variable-sharing are modeled in a common framework. The basic notions of IMDS are **communication items** and **atomic actions**, and its basic benefit is that a system can be decomposed in various ways into **components**.

Communication items are partitioned among system nodes. There are two types of items, **consumable passed items** (representing messages, remote procedure calls etc.) and **reusable stored items** (representing a state of a system **node**). Stored items are pairs (**location**, **resource**), where location identifies the node in a system and resource represents its internal state (validation over variables). Passed items are triples (**destination**, **service**, **tag**), where destination identifies a node to which a message or call is directed, service identifies 'entry point' such as procedure and tag identifies a 'matter' in which a call is issued. For example, a client node may issue two calls to server: 'tell me distance between A and B' and 'tell me distance between C and D'. The two calls may be pending at the same time and different tags distinguish them.

An atomic **action** processes a passed item and a stored item on a node and creates some passed items and stored items. More precisely, an action consumes two input items: a stored item and a passed one with location and destination equal. It produces a number of output items: obligatory stored item with the same location (called **continuation stored item** – a new value of a node) and optional passed item with the same tag (called **continuation passed item**). If there is no continuation passed item then the sequence of operations in given matter (identified by a tag) stops. An action may produce stored items with fresh location (which did not occur in the system yet): this produces a new node. An action may also produce passed items with fresh tag: this branches new sequences.

A **quota** of a system is a set of all its communication items. In the operation of system **decomposition**, a component is specified by its own quota of items, a subset of system quota.. The sets of passed item quotas of composable components are pair-wise disjoint; their sets of stored items can overleap.

A **system configuration** is a subset of its quota. In a system decomposition, a **component configuration** is the intersection of the system configuration and the component quota. An action is prepared if its input items are present in current system configuration (we say that these items are pending). Firing an action transforms a system configuration into a new system configuration, in which input items on the action are removed and output items are added. An **initial configuration** of a system contains at most one stored item with any location and at most one passed item with any tag.

A system action is seen as **a multi-handshake** of sub-actions of those components whose quotas contain some items processed or created by the action. We call such components **participants** of the action. We say that a participant component **delivers** some items processed in the action or **takes** over some items created by it.

## 1.2 Communication in a System

Two components, such that one delivers a passed item and the other one a stored item processed by an action, are said to communicate in a **synchronous** way in the action. On the other hand, if one component delivers some processed items and there is a component which takes over created item, they communicate in an **asynchronous** way in the action. If the item taken over is a passed item then it is a **message passing**, otherwise it is a **variable sharing**. If a delivered item and a retrieved item belong to a quota of a single component then it is a communication internal to the component; Otherwise it is an external communication (or inter-component communication). It follows that in various decompositions, some communication forms are hidden as internal communication and others are exposed as external communication.

In the whole spectrum of possible system decompositions, we identify two major decompositions into asynchronous sequential processes and define the notions of inter- and intra-component communication. In the first decomposition, components execute their actions on various system nodes and hide message passing as intra-component communication and expose variable sharing as inter-component communication. We call such components traveler processes and the decomposition the shared-variable decomposition. A traveler's quota is a set of all passed items with equal tag, and all stored items. Travelers communicate via shared stored items (interpreted as common variables) created on output of actions and available for other processes as input items of their actions. It is just opposite in the other decomposition where each component executes its actions on a single node; the component hide variable sharing as intra-component communication and expose message passing as inter-component communication. We call such components resident processes and the decomposition the message-passing decomposition. A resident's quota is a set of all items with given node as location or destination. Residents communicate via passed items (interpreted as messages) created on output of actions and directed to nodes as destinations of items.

We call the transposition of the roles of passing and sharing as inter- and intra-process communication in the two decompositions **communication dualism.** We will show that the two decompositions are **canonical decompositions**.

In the light of this property, communication in distributed systems can be seen as a two-dimensional phenomenon with passing being its spatial dimension and sharing its temporal dimension. Communication dualism shows that shared memory is not a necessary substrate upon which sharing can exist and that message passing is not the only direct communication in distributed memory. Thus, any distributed system can be modeled as a composition of message-passing asynchronous processes or as a composition of variable-sharing asynchronous processes. This follows the famous Lauer-Needham communication dualism hypothesis [16].

The integration of communication and formalization of communication dualism in our model is possible due to a generalization of the process concept. The crucial point is that the IMDS model contains both classical description using processes resident on system nodes, and novel concept of traveler processes as well. In fact, a kind of traveler processes can be found in shared-memory systems, too. The conviction that processes in shared memory reside on processors is only an illusion caused by the fact that most of shared memory communication functions are implemented in hardware. As a result, a shared-memory system is often considered as a single (tightly coupled) object, see e.g.: [17]. At an abstraction level, where each shared memory cell and each processor are distinct objects, a process executes its actions at the processor and at memory cells. In this sense it travels among the processor and memory cells [3,5].

## 1.3  Related work

In most of the classical models of distributed systems and in computational models of process calculi, a system is modeled as a composition of processes communicating in a predefined way. In many models this is synchronous message passing. This approach leads to a concise model, but assumes a primary (an absolute) communication on top of which other communication forms are defined. In such models, neither an communication integration can be modeled nor communication dualism can be formalized.

Hoare's CSP [14] and Milner's CCS [18,19] introduce handshake as a fundamental concept in modeling communication. Its outstanding feature is elimination of differences between agents and their media. However, both the models abstract away distribution of a system. Moreover, handshake does not model asynchronous communication.

In the π calculus [20,21], an extension of CCS to distributed setting, handshake is reinterpreted as a rendezvous at the ends of a synchronous channel. In this way, the difference between processes and their media is re-established, there is no locality of actions, and both multiple receiver channels and the parallel waiting for input and output from various channels assume a global state access in a system action. The last assumption is in contradiction with the fundamental property of system distribution, i.e. that only local state is accessible in an atomic system action. The access to global state is non-implementable in distributed environment.

In most of classical message passing languages, the global state access is avoided by imposing a restriction that a process can wait in a parallel way for channel input operations only (see for example, *select* in Ada, *ALT* in Occam). However, this restriction, in our opinion, is only a by-pass solution. In its channel output operation, a process can be blocked (and messages pending in other channels are not received) until its counterpart is ready with its input operation. We call this blocking synchronization overhead.

In other models [22,23], the synchronization overhead is eliminated by replacing a sequential process at each distributed site, with a set of parallel threads. In distributed versions of such models, threads communicate by channels and sites communicate by special primitives for passing messages [24,25].

In our opinion, in a radical solution, passing channels should be eliminated and passing and sharing should be modeled as peer communications in terms of consumable passed items and reusable shared items.

In the join calculus [10,11,12], channels are removed and locality is established, but passed and shared items are not distinguished. This is why, consumability of passed items and reusability of stored items have to be established at application level and recognized (e.g.: for optimization purposes) at compilation stage [13]. In an effect, in the join calculus, shared-variable communication is not modeled directly [1].

In this paper, we formalize a behavior of a modeled system, an operation of a system decomposition and the two canonical decompositions of a system. On this ground, we introduce the concept of communication dualism. In related work, we verify the statements of some classical distributed problems in the context of communication dualism and discus the way shared-variable communication can be introduced to the computational models of process calculi.

## 2 Behavior of a System

We model a system behavior using a version of a labeled transition system whose configurations, actions and transitions are expressed in terms of communication items and whose outstanding feature is an operation of system decomposition.

### 2.1 Set Theory Specification

A set theory specification of a system is a tuple of five pair-wise disjoined sets:
- *LAB={l, l', …}* – **labels** identifying system **nodes**
- *SER={se, se', …}* – **services** called on nodes
- *RES={re, re', …}* – **resources** representing states of nodes
- *TAG={t, t', … }* – **tags** representing "matters" or "control flow identifiers" a system deals with
- *ACT={$\lambda$, $\lambda$', … }* – **system actions** determining a progress in a system

A **passed item** is a triple $p = t.l.se \in TAG \times LAB \times SER = PAS$ containing tag *t*, destination *l* and service *se*. The destination *l* is the label of the node the item *p* is directed to. The service distinguishes among various procedures/entry points/facilities offered by the destination node. Passed items with the same destination and the same service are identified with unique tags. For every *p*, *tag(p)* is the tag of *p* and *destination(p)* the destination of *p*.

A **stored item** is a pair $s = l.re \in LAB \times RES = STO$ containing its location *l* and resource *re*. The location of a stored item *s* is the label of the node the item *s* is stored on. The resource identified the state of the node, i.e. it differentiates between all possible valuations over internal variables of the node. For every *s, location(s)* is the location of *s*.

A **communication item** is simply a stored or passed item. For every label *l*, $PAS_l$ is the set of all passed items with the destination *l* and $STO_l$ is the set of all stored items with the location *l*. Similarly, for every tag *t*, $PAS_t$ is the set of all passed items with the tag *t*.

We denote as *ITE* the set of all communication items and let *i, i'*, … range over *ITE*. We extend the definitions of *tag()*, *destination()* and *location()* to any *set* $I \subseteq ITE$:
- *tags(I)* = $\cup_{i \in I \cap PAS}$ *{tag(i)}*,
- *destinations(I)* = $\cup_{i \in I \cap PAS}$ *{destination(i)}*,
- *locations(I)* = $\cup_{i \in I \cap STO}$ *{location(i)}*,

We denote the set of all labels in *I* as:
- *labels(I)* = $\cup_{i \in I \cap PAS}$ *{(destination(i)}* $\cup$ $\cup_{i \in I \cap STO}$ *{location(i)})*

For any $I \subseteq ITE$, **H**(I) is the set of all subsets such that, in each subset, for every *t*, there is at most one passed item with the tag *t* and, for each *l*, at most one stored item with the location *l*. For any $P \subseteq PAS$ and any $S \subseteq STO$, **D**(P, S) = {{p, s} | $p \in P, s \in S$}.

(2.1.1) The set of **actions** *ACT* is a function: *ACT:* **D**(PAS, STO) $\Rightarrow$ **H**(ITE); such that
    (a) $\lambda = \{<\{p,s\}, CI>\} \in ACT$
    (b) $\forall_{\lambda \in ACT}$ *destination(p) = location(s)*,    // executed on single node
    (c) $\forall_{\lambda \in ACT}$ $\exists_{cs \in (CI \cap STO)}$ *location(s) = location(cs)*,
                                       // obligatory continuation stored item
    (d) $\forall_{\lambda \in ACT}$ *tags(CI) – ({tag(p)} $\cap$ tags($\gamma$)) = $\varnothing$*    // new tags in CI are fresh;
    (e) $\forall_{\lambda \in ACT}$ *(labels(CI)- ({location(s)} $\cap$ labels($\gamma$)) = $\varnothing$*
                                       //new labels in CI are fresh;

For any action $\lambda = <\{p, s\}, CI>$, the stored item $cs \in CI$, $location(cs)=location(s)$ is called the continuation stored item of $\lambda$; if there is a passed item $cp \in CI$, $tag(cp) = tag(p)$ then $cp$ is called the continuation passed item of $\lambda$.

(2.1.2) A set $\gamma \subseteq H(ITE)$ is called a **system configuration** iff:
- $labels(\gamma) = locations(\gamma)$.

//all passed items are directed to nodes in the configuration

An action $\lambda = <\{p,s\}, CI>$ is **prepared** in a system configuration $\gamma$, iff $\{p,s\} \subseteq \gamma$. An action transforms a system configuration $\gamma$ into a new system configuration $\gamma'$ such that $\{p,s\} \subseteq \gamma$, $CI \subseteq \gamma'$ and $\gamma - \{p,s\} = \gamma' - CI$ (see next section – LTS). Such a transformation we call **firing** the action. As $destination(p)=location(s)$, we may say that the action is "executed on the node".

A Petri net interpretation of an example action $\lambda = <\{p,s\},\{p',s'\}>$ is presented in Fig. 1. Tokens in input places represent items present in a system configuration (see next section – LTS).

## 2.2 Labeled Transition System

The behavior of a system is defined as a labeled transition system, $LTS = <\Gamma, \Lambda, \Theta>$, such that:

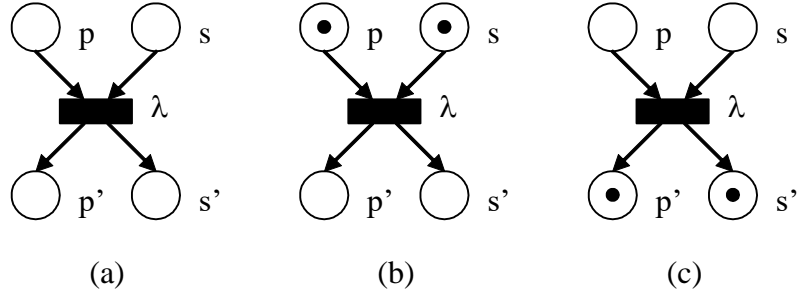

(a)    (b)    (c)

**Fig. 1** Petri net interpretation of IMDS action (a), prepared action (b) and action after its firing (c)

- $\Gamma$ is a set of **system states** - all **maximal system configurations** $\gamma \in H(ITE)$, (maximal denotes that $\forall_{\gamma,\gamma' \subseteq H(ITE), \gamma \neq \gamma'} \gamma \not\subset \gamma'$ and $\gamma' \not\subset \gamma$),
- $\Lambda$ is a set of **system labels** – all system actions $\lambda = <\{p,s\}, CI> \in ACT$,
- $\Theta$ is a set of **system transitions** $\tau = <\gamma, \lambda, \gamma'>$; $\lambda = <\{p,s\}, CI>$; $\{p,s\} \subseteq \gamma$; $\gamma' = (\gamma - \{p,s\}) \cup CI$.

The semantic issues of *LTS* are:
1. Prepared actions executed on distinct nodes do not depend on each other:
    - input stored items are distinct, because they have different locations;
    - input passed items are distinct, because they have different destinations;
    - continuation stored items are distinct, because they have different locations;
    - continuation passed items (if present) are distinct, because they have different tags (two items with equal tag cannot exist in a system configuration, see 2.1.2);
    - output items other than continuation items are fresh and different.

Therefore, any concurrency model between actions on distinct nodes may be assumed:
- interleaving (one action fired at a time);
- maximum concurrency (all prepared actions fired simultaneously);
- intermediate model: any positive number of prepared actions may be fired (between one and all prepared actions).

A chosen model of concurrency may influence system runs (paths of fired actions).
2. Prepared actions executed on a single node are always in conflict, as they use equal input stored item. Only one of them may be fired at a time. But the action may be chosen from prepared ones arbitrary: nondeterministic choice is the best solution ensuring fairness of the system.

The above semantic issues reflect the intuition on distributed systems: sequential (but potentially nondeterministic) execution of operations on a single node, freedom of choice of concurrency model on distinct nodes.

In Petri net interpretation, a connecting of actions (transitions) by items (places) is shown in Fig. 2. There are three actions $\lambda_1 = <\{p_1,s_1\},\{p_1',s_1'\}>$, $\lambda_2 = <\{p_2,s_2\},\{p_2',s_2'\}>$, $\lambda_3 = <\{p_2',s_1'\},\{p_3',s_3'\}>$. Firing rules of Petri net transition reflect strictly firing rules of an action: input items must be pending, i.e. input places must contain tokens. Tokens are placed in places representing the items in initial configuration of the system.

For system analysis purpose, the Petri net should be safe – after certain number of transitions:
- the number of nodes need not grow (stored items with fresh location need not occur), as the existence of the node is represented as a token in one of places addressed to stored items with location identifying the node;
- the number of tags need not grow (the number of passed items with fresh tags should be equal to the passed of tags "consumed": not occurring in continuation passed items).

## 3 Decomposition of System

An operation of a system decomposition into components is among the main features of IMDS. Intuitively, a component is a projection of the system onto a subset of communication items, called its quota. In various decompositions, components can communicate in synchronous way and/or asynchronous passing/sharing way.

### 3.1 Component and a System Decomposition

A component $W(Q)$ in a system is specified by a subset $Q \subseteq ITE$ called the **component quota**. The set of behaviors of a component $W(Q)$ is a labeled transition system, $<C, A, TR>$ such that:

(3.1.1) (a) set of **component states** of $W(Q)$ such that for any system configuration $\gamma$, $c = \gamma \cap Q \in C$ is a **component configuration** of $W(Q)$ in $\gamma$;

(b) $A$ is a set of **component labels** of $W(Q)$ such that for any system action $\lambda = <\{p,s\}, CI>$, the pair $\alpha = <\{p,s\} \cap Q, CI \cap Q> \in A$ is a **component action share** of $W(Q)$ in $\lambda$;

(c) $TR$ is a set of **component transitions** $tr = <c, \alpha, c'>$ in $W(Q)$; a component action share $\alpha = <\{p,s\} \cap Q, CI \cap Q>$; $\{p,s\} \cap Q \subseteq c$ makes a component transition from a component configuration $c \in Q$ to a component configuration $c' \in Q$, such that: $c' = (c - (\{p,s\} \cap Q)) \cup (CI \cap Q)$.

We say that a component action share $\alpha = <\{p,s\} \cap Q, CI \cap Q>$ is prepared in a component configuration $c$ of $W(Q)$, if $\{p,s\} \cap Q \subseteq c$;

For any $Q$, $QP = Q \cap PAS$ is called the **passed item quota** of $W(Q)$ and $QS = Q \cap STO$ is called the **stored item quota** of $W(Q)$. Components of a system are **composable**, if their passed item quotas are pairwise disjoint (their stored item quotas can overlap). For any $Q_1, Q_2 \in ITE$, if components $W(Q_1)$ and $W(Q_2)$ in a system are composable, then $W(Q_1 \cup Q_2)$ is also a component in the system.

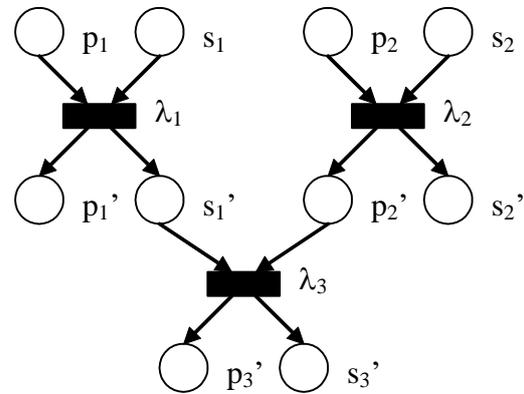

**Fig. 2** Petri net interpretation of IMDS system

(3.1.1) Any set $D = \{W(Q_k = QP_k \cup QS_k) \mid k \in K\}$ is a **system decomposition**, iff $\{QP_k \mid k \in K\}$ is a partition of *PAS* and $\cup_k QS_k = STO$.

We let *D* range over the set of all decompositions. For every decomposition $D = \{W(Q_k) \mid k \in K\}$, $QD(D) = \{Q_k \mid k \in K\}$ is the **quota determinant** of *D*.

## 3.2 Communication in System Decomposition

In any decomposition *D*, each system action $\lambda$ is seen as a parallel composition of non empty component shares. We call it the **multi-handshake** of $\lambda$ in *D*.

(3.2.1) In a decomposition *D* containing components $W(Q_1)$ and $W(Q_2)$, we define the following **communication forms** in the multi-handshake of any system action $\lambda = <\{p,s\}, CI>$:

(a) $W(Q_1)$ and $W(Q_2)$ communicate in $\lambda$ in a **synchronous** way, iff $\{p,s\} \cap Q_1 = \{p\}$ and $\{p,s\} \cap Q_2 = \{s\}$;
//one component delivers passed item and the other one delivers stored item

(b) $W(Q_1)$ and $W(Q_2)$ communicate in $\lambda$ in an **asynchronous** way by item *i*, iff $\{p,s\} \cap Q_1 \neq \varnothing$ and $\exists_{i \in Q_2} i \in (CI \cap Q_2)$, in particular:

(c) $W(Q_1)$ and $W(Q_2)$ in $\lambda$ **communicate by passing item *i***, iff $i \in PAS$,

(d) $W(Q_1)$ and $W(Q_2)$ in $\lambda$ **communicate by sharing item *i***, iff $i \in STO$.
//one component delivers an input item and the other one takes an output item

Communication between $W(Q_1)$ and $W(Q_2)$, such that $Q_1 \neq Q_2$ is an **external communication** in a decomposition *D*; if $Q_1 = Q_2$ then it is **internal communication** in the in decomposition *D*.

# 4 Canonical Decompositions

There are many ways a system may be decomposed in. For example, there are decompositions:

- to elementary processes (single item processes),
- to synchronous sequential processes (messengers with tags and custodians with locations),
- semi-synchronous decomposition corresponding to the asynchronous $\pi$ calculi [2,15], etc.

All of them are described in detail in [7].

In this paper we study two special decompositions of a system into asynchronous processes and define the inter-process communication concept. In one of the decompositions, inter-process communication is message passing and, in the other, variable sharing. We call them canonical decompositions and their components – canonical processes. Brought into a common framework of IMDS they manifest the communication dualism property.

## 4.1 Asynchronous Process

A component share $\alpha = <\{p,s\} \cap Q, CI \cap Q>$, such that $\{p,s\} \subseteq Q$ is called a **processing action share**. Consider a decomposition where each system action has a processing action share. It means that, in such a decomposition, synchronous communication is hidden as internal communication of its components and the components communicate in an asynchronous way only. We call them **asynchronous processes**.

It follows that, in each component configuration of any asynchronous process $W(Q)$, each pending passed item *p* has a matching current stored item s to be processed; otherwise $W(Q)$ would need another process $W(Q')$ with a current stored item matching *p*. In this case $W(Q)$ and $W(Q')$ would process *p* synchronously. We also assume that, in the quota of an asynchronous process, there are no superfluous stored items which are never processed by the process. More precisely, (4.1.1) For any quota $Q$, $W(Q)$ is an asynchronous process, iff, for each of its component configurations $c$, $destinations(c) \subseteq locations(c)$ and for each stored item $s$ in $Q$, there exists a configuration $c$, such that $location(s) \in destinations(c)$.

This definition implies the following condition for the quota of an asynchronous process:

**Lemma 4.1. The quota of an asynchronous process**

For any quota $Q$, the component $W(Q)$ is an asynchronous process, iff

$Q \cap STO = \cup_{p \in (Q \cap PAS)} STO_{destination(p)}$.

Lemma 4.1 allows an asynchronous process to be specified by its passed item quota only: the subsystem $PR(QP) = W(QP \cup (\cup_{p \in QP} STO_{destination(p)}))$ is called an asynchronous process with the passed item quota $QP$.

## 4.2 Asynchronous Sequential Process

In the classical models of shared-variable and message-passing languages, an **asynchronous sequential process** is the most commonly used type of a process. It performs its processing action shares one at a time. In modeling sequentiality in IMDS, we make use of the fact that all actions on a node are performed in a mutually exclusive way and we postulate that in any component configuration $c$ of a sequential asynchronous process, all processing shares prepared in $c$, if any, are performed on single node only. More precisely,

(4.2.1) For any passed item quota $QP$, $PR(QP)$ is an asynchronous sequential process, iff for any of its component configuration $c$, there exists a label $l$, such that for any pending passed item $p \in c$, $destination(p)=l$.

The following Lemma states a property of an asynchronous sequential process quota.

**Lemma 4.2. The quota of an asynchronous sequential process**

For any $QP$, the process $PR(QP)$ is an asynchronous sequential process, iff $\exists_{t \in TAG}[QP \subseteq PAS_t]$ or $\exists_{l \in LAB}[QP \subseteq PAS_l]$.

The set $SP$ of all asynchronous sequential processes can be partitioned in two ways: into classes of **tag-oriented processes**, $\forall_{t \in TAG} [TP(t) = \{PR(QP) \mid QP \subseteq PAS_t\}]$ and into classes of **label-oriented processes** (or node-oriented processes), $\forall_{l \in LAB} [LP(l) = \{PR(QP) \mid QP \subset PAS_l\}]$, in such a way that:

- $\{TP(t) \mid t \in TAG\}$ is a partition of $SP$ and
- $\{LP(l) \mid l \in LAB\}$ is a partition of $SP$.

In each equivalence class, the processes are ordered. For any $t \in TAG$, the binary relation, $<$, on the set $TP(t)$, such that, $PR(QP) < PR(QP')$, iff $QP \subset QP'$ is a partial order relation and $PR(PAS_t)$ is the maximum element in the set $TP(t)$. For any $l \in LAB$, the binary relation, $<$, on the set $LP(l)$, such that $PR(QP) < PR(QP')$, iff $QP \subset QP'$ is a partial order relation and $PR(PAS_l)$ is the maximum element in the set $LP(l)$.

## 4.3 Canonical Process

The maximum process in each equivalence class of tag-oriented processes or label-oriented processes is called a canonical process. The tag-oriented canonical process is called the **traveler** with tag $t$ and the label-oriented canonical process is called the **resident** process on the node $l$. Formally,

(4.3.1) For any $t$, $TR_t = PR(PAS_t)$ is called the traveler process with the tag $t$. For any $l$, $RE_l = PR(PAS_l)$ is called the resident process with the label $l$.

The processing action shares of any traveler can be performed on various nodes, while all the processing shares of a resident are performed on a single node. Lemma 4.1 implies the following full quotas for canonical processes:

(a) For any $t \in TAG$, $TR_t = W(PAS_t \cup STO)$,
(b) For any $l \in LAB$, $RE_l = W(PAS_l \cup STO_l)$.

Using the facts that *{PAS_t | t ∈ TAG}* and *{PAS_l | l ∈ LAB}* are partitions of *PAS* and that $\cup_{l \in LAB} STO_l = STO$ it is easy to prove the central theorem in our model:

**Theorem 4.3. The canonical decompositions of a system**

(a) *TD = {TR_t | t ∈ TAG}* is a system decomposition. It is called the **traveler decomposition**.

(b) *RD = {RE_l | l ∈ LAB}* is a system decomposition. It is called the **resident decomposition**.

## 4.4 Canonical processes in Petri net interpretation.

Processes in Petri nets are defined as sequences of fired actions. For some decompositions it is hard to extract such processes, for example in a case of synchronous messenger-custodian decomposition any action belong to two processes (a messenger delivering a passed item an a custodian delivering a stored item).

Fortunately, in asynchronous process model an action is executed entirely in a process. A canonical process quota is determined by a tag or by a label. In a case of resident decomposition (Fig. 3), a process starts statically in initial marking if there is a token in an item (place) with its label (a), or dynamically in an action starting a new label (b). The run of the process follows the label (c), therefore continuation stored item determines the process control flow. In a case of traveler decomposition (Fig. 4), a process starts statically in initial marking if there is a token in an item (place) with its tag (a), or dynamically in an action starting a new tag (b). The run of the process follows the tag (c), therefore continuation passed item determines the process control flow. If there is no continuation passed item, the process terminates (d).

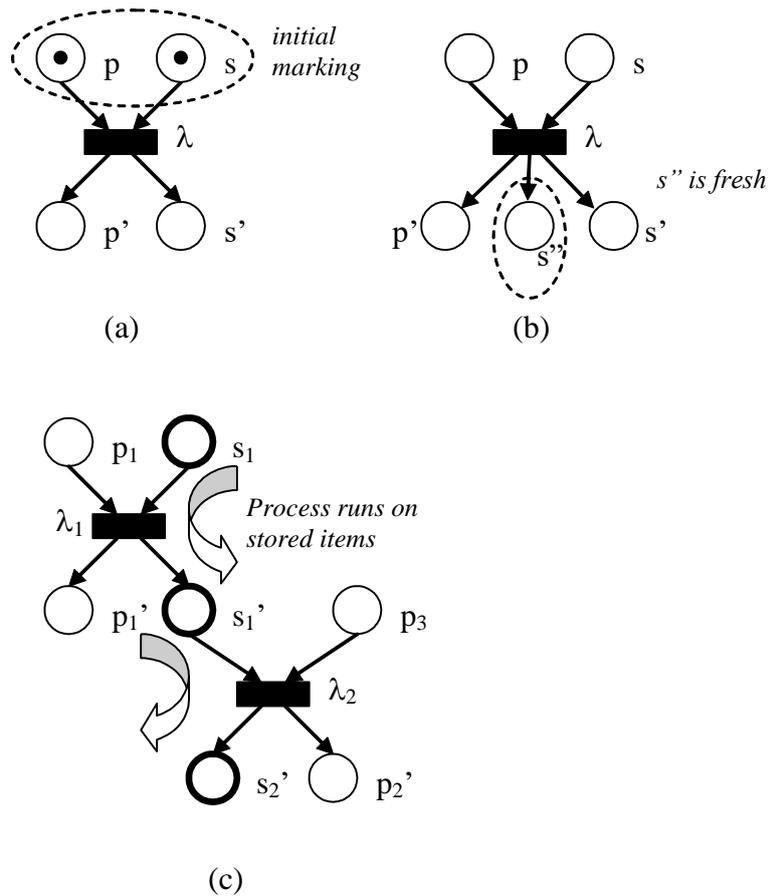

**Fig. 3** Petri net interpretation of resident process: static start (a), dynamic start (b) and control flow (c)

Automatic extraction of processes from a system may be simplified by coloring of tokens. We define two classes of colors: $PCOL=\{ct_1,ct_2,...\}$ for tokens falling into passed items, $SCOL=\{cl_1,cl_2,...\}$ for tokens falling into stored items. If tags disappear due to lack of continuation passed items and new (fresh) tags originate, colors of these tags should be reused.

A token falling into a place representing a passed item $p=t_i.l.se$ gets color $ct_i$, A token falling into a place representing a stored item $s=l_i.re$ gets color $cl_i$. Note that a token moved from a passed item to a continuation passed item (if any) preserves color as the tag in both items is the same. A token moved from a stored item to a continuation stored item preserves color as the location in both items is the same. Tokens in fresh items get fresh colors (or reused ones). Therefore, colors of tokens moved in actions (transitions) from input items to continuation items of actions form processes: travelers (*PCOL*) or residents (*SCOL*) respectively. New (or reused) colors form new processes, lost colors terminate processes.

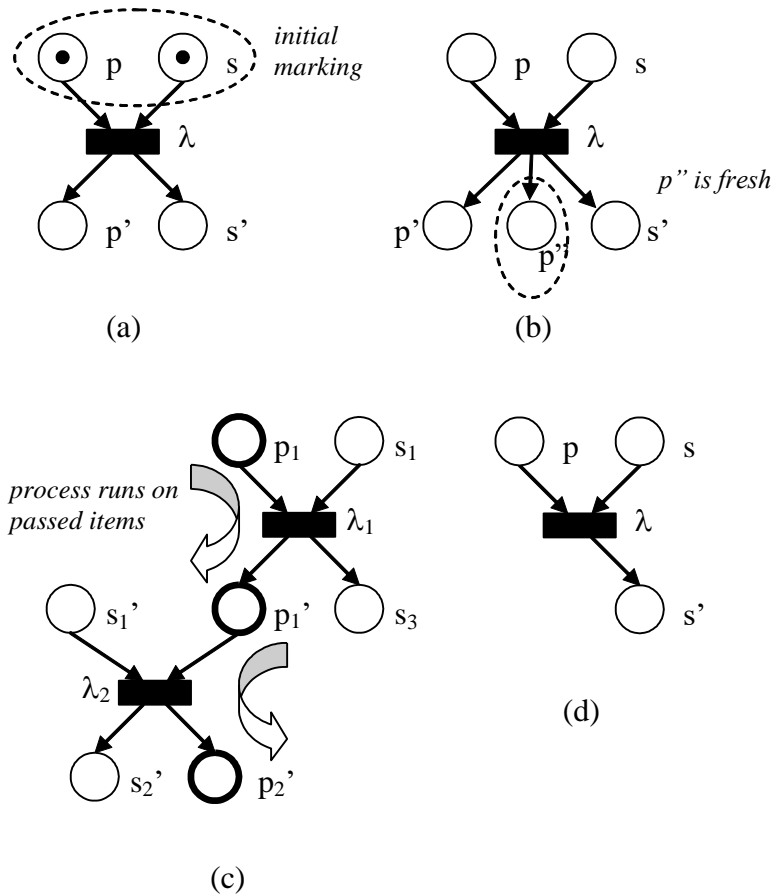

**Fig. 4** Petri net interpretation of traveler process: static start (a), dynamic start (b), control flow (c), and stop (d)

## 5 Future work

The IMDS formalism allows a designer to express features which are hard or impossible to be expressed in other formalisms [8]. For example:

- distributed mutual exclusion,
- distributed termination (and distinguishing it from deadlock) [4],
- communication deadlock [6],
- deadlock over resources (and distinguishing it from communication deadlock),
- partial deadlock (when only a part of all processes fall into deadlock).

Integration of IMDS with a model checking formalism (precisely: expressing items and actions in terms of state space for verification and expressing features like termination or deadlock as verification formulas) will allow to verify distributed system against these features and to generate counterexamples explaining the reasons of errors found [9].

## 6 Conclusion

Integration of various communication forms in a common framework reveals a significant relationship between communication forms in distributed systems, namely communication dualism.

Communication dualism allows realizing that passing and sharing are peer communication forms, that communication is a two-dimensional phenomenon and that passing is its spatial and sharing is its temporal manifestation.

The significance of communication dualism can be observed in both theoretical and experimental fields of distributed systems. First of all, it shows that inter-process communication in a system does not depend on the substrate upon which the system is built, but on the way the system is decomposed into processes. It implies that message passing is not the only direct communication form in distributed-memory systems and that some properties of the system may be defined in terms of shared communication items and others in terms of both passed items. A straight consequence of communication dualism is that equational theories of process calculi may and should be extended to passing and sharing versions of a system.

## Acknowledgment

We wish to thank Jerzy Mieścicki for his help.

# Appendix (contains proofs of some lemmas)

## Lemma 4.1. The quota of an asynchronous process

For any quota $Q$, the subsystem $W(Q)$ is an asynchronous process, iff
$Q \cap STO = \cup_{p \in (Q \cap PAS)} STO_{destination(p)}$.

**Proof**: Using (4.1.1), we need to prove that

(1) $\forall_{Q \subseteq ITE}[ \forall_{c \subseteq Q}[destinations(c) \subseteq locations(c)] \equiv$
$$\equiv (Q \cap STO) \supseteq \cup_{p \in (Q \cap PAS)} STO_{destination(p)}]$$

and

(2) $\forall_{Q \subseteq ITE} [ \forall_{s \in (Q \cap STO)} \exists_{c \subseteq Q}[locations(\{s\}) \subseteq destinations(c)] \equiv$
$$\equiv (Q \cap STO) \subseteq \cup_{p \in (Q \cap PAS)} STO_{destination(p)}].$$

Using $c = \gamma \cap Q$ we transform (1) and (2) into equivalent formulas (1') and (2').

(1') $\forall_{Q \subseteq ITE} [ \forall_{\gamma \subseteq ITE} [destinations(\gamma \cap Q) \subseteq locations(\gamma \cap Q)] \equiv$
$$\equiv (Q \cap STO) \supseteq \cup_{p \in (Q \cap PAS)} STO_{destination(p)}]$$

and

(2') $\forall_{Q \subseteq ITE} [ \forall_{s \in (Q \cap STO)} \exists_{\gamma \subseteq ITE} [locations(\{s\}) \subseteq destinations(\gamma \cap Q)] \equiv$
$$\equiv (Q \cap STO) \subseteq \cup_{p \in (Q \cap PAS)} STO_{destination(p)}].$$

We prove both (1') and (2') by proving the contra-positions of their component implications.

(1'.a) $\forall_{Q \subseteq ITE} [ \cup_{p \in (Q \cap PAS)} STO_{destination(p)} \not\subseteq (Q \cap STO)] \Rightarrow$

$\forall_{Q \subseteq ITE} [\exists_{p \in PAS, s \in STO}[s \in \cup_{q \in (Q \cup PAS)} STO_{destination(q)}, s \notin (Q \cap STO), p \in (Q \cap PAS)]] \Rightarrow$

$\forall_{Q \subseteq ITE} [\exists_{p \in PAS, s \in STO, \gamma \subseteq ITE} [\ \gamma = \{p,s\},$
$\qquad destinations(\gamma \cap Q) = \{destination(p)\} = \{location(s)\},$
$\qquad locations(\gamma \cap Q) = loccations(\{\}) = \varnothing]] \Rightarrow$

$\forall_{Q \subseteq ITE} [\exists_{\gamma \subseteq ITE} [destinations(\gamma \cap Q) \not\subseteq locations(\gamma \cap Q)]]$.

(1'.b) $\forall_{Q \subseteq ITE} [\exists_{\gamma \subseteq ITE} [destinations(\gamma \cap Q) \not\subseteq locations(\gamma \cap Q)] \Rightarrow$

$\forall_{Q \subseteq ITE} [\exists_{\gamma \subseteq ITE} [\ \forall_{p \in PAS, s \in STO} [p \in (\gamma \cap Q), s \in \gamma, s \notin Q, destination(p) = location(s)]]] \Rightarrow$

$\forall_{Q \subseteq ITE} [\exists_{\gamma \subseteq ITE} [\ \forall_{p \in PAS, s \in STO} [s \in STO_{destination(p)}, s \notin (Q \cap STO)]]] \Rightarrow$

$\forall_{Q \subseteq ITE} [\exists_{\gamma \subseteq ITE} [\cup_{p \in (Q \cap PAS)} STO_{destination(p)} \not\subseteq (Q \cap STO)]]$.

(1'.a) and (1'.b) proves (1).

(2'.a) $\forall_{Q \subseteq ITE} \exists_{s \in STO} \forall_{\gamma \subseteq ITE} [s \in (Q \cap STO), location(s) \notin destinations(\gamma \cap Q)]] \Rightarrow$

$\forall_{Q \subseteq ITE} \exists_{s \in STO} [s \in (Q \cap STO), \forall_{\gamma \subseteq ITE} [location(s) \notin destinations(\gamma \cap Q)]] \Rightarrow$

$\forall_{Q \subseteq ITE} \exists_{s \in STO} [s \in (Q \cap STO), locations(s) \notin destinations(Q \cap PAS)] \Rightarrow$

$\forall_{Q \subseteq ITE} \exists_{s \in STO} [s \in (Q \cap STO), s \notin (\cup_{p \in (Q \cap PAS)} STO_{destination(p)})]$.

(2'.b) $\forall_{Q \subseteq ITE} [(Q \cap STO) \not\subseteq \cup_{p \in (Q \cap PAS)} STO_{destination(p)}] \Rightarrow$

$\forall_{Q \subseteq ITE} \exists_{s \in STO} [s \in (Q \cap STO), \forall_{p \in (Q \cap PAS)} [location(s) \notin destinations(p)]] \Rightarrow$

$\forall_{Q \subseteq ITE} \exists_{s \in STO} \forall_{\gamma \subseteq ITE} [s \in (Q \cap STO), location(s) \notin destinations(\gamma \cap Q)] \Rightarrow$

$\forall_{Q \subseteq ITE} \exists_{s \in STO} \forall_{\gamma \subseteq ITE} [s \in (Q \cap STO), locations(\{s\}) \notin destinations(\gamma \cap Q)]$

(2'.a) and (2'.b) prove (2), which together with (1) proves the Lemma. □

### Lemma 4.2. The quota of a sequential asynchronous process

For any *QP*, the process *PR(QP)* is a sequential process, iff
$\exists_{t \in TAG} [QP \subseteq PAS_t]$ or $\exists_{l \in LAB} [QP \subseteq PAS_l]$.

**Proof:** Using (4.1.1) we need to prove the equivalence

$\forall_{QP \subseteq ITE} [\ \forall_{c \subseteq ITE} \exists_{l \in LAB} [destinations(c) = \{l\}] \equiv$
$\qquad \equiv \exists_{t \in TAG} [QP \subseteq PAS_t]$ or $\exists_{l \in LAB} [QP \subseteq PAS_l]]$.

1. $\forall_{QP \subseteq ITE} [\exists_{t \in TAG} [QP \subseteq PAS_t]$ or $\exists_{l \in LAB} [QP \subseteq PAS_l] \equiv$

2. $\equiv \exists_{t \in TAG} \forall_{\gamma \subseteq ITE} [(\gamma \cap QP) \subseteq PAS_t]$ or $\exists_{l \in LAB} \forall_{\gamma \subseteq ITE} [(\gamma \cap QS) \subseteq PAS_l] \equiv$

3. $\equiv \forall_{\gamma \subseteq ITE} \exists_{t \in TAG} [(\gamma \cap QP) \subseteq PAS_t]$ or $\forall_{\gamma \subseteq ITE} \exists_{l \in LAB} [(\gamma \cap QS) \subseteq PAS_l] \equiv$

4. $\equiv \forall_{\gamma \subseteq ITE} [\ \forall_{p,p' \in PAS} [p,p' \in (\gamma \cap QP), tag(p) = tag(p')]$ or
$\qquad\qquad\qquad \forall_{\gamma \subseteq ITE} \exists_{l \in LAB} [(\gamma \cap QP) \subseteq PAS_l] \equiv$

5. $\equiv \forall_{\gamma \subseteq ITE} [\ \forall_{p,p' \in PAS} [p,p' \in (\gamma \cap QP), p=p']$ or $\forall_{\gamma \subseteq ITE} \exists_{l \in LAB} [(\gamma \cap QP) \subseteq PAS_l] \equiv$

6. $\equiv \forall_{\gamma \subseteq ITE} \exists_{l \in LAB} [(\gamma \cap QP) \subseteq PAS_l]$ or $\forall_{\gamma \subseteq ITE} \exists_{l \in LAB} [(\gamma \cap QP) \subseteq PAS_l] \equiv$

7. $\quad \equiv \forall_{\gamma \subseteq ITE} \exists_{l \in LAB} [(\gamma \cap QP) \subseteq PAS_l]].$

In the pass from 4. to 5. we make use of (3.2.1.a) which implies:

$\forall_{\gamma \subseteq ITE} \forall_{p,p' \in PAS} [p,p' \in \gamma, tag(p) = tag(p') \Rightarrow p = p'].$ □